\documentclass[fleqn,12pt,twoside]{article}
\usepackage{espcrc1}

\usepackage{graphicx}
\usepackage[figuresright]{rotating}

\newcommand{\AmS}{{\protect\the\textfont2
  A\kern-.1667em\lower.5ex\hbox{M}\kern-.125emS}}

\title{Mass and Isospin Effects in Multifragmentation}

\author{C.~Sfienti\address[GSI]{Gesellschaft f{\"u}r Schwerionenforschung, 
D-64291 Darmstadt, Germany}\address[INFNCT]{Dipartimento di Fisica
dell'Universit\'{a} and LNS-INFN, I-95126 Catania, Italy},
P.~Adrich\addressmark[GSI], 
T.~Aumann\addressmark[GSI], 
C.O.~Bacri\address[FRA1]{Institut de Physique Nucl{\'e}aire, IN2P3-CNRS 
et Universit{\'e}, F-91406 Orsay, France}, 
T.~Barczyk\address[POL2]{M.~Smoluchowski Institute of Physics, Jagiellonian
Univ., Pl-30059 Krak{\'o}w, Poland}, 
R.~Bassini\address[INFNMI]{Istituto di Scienze Fisiche, Universit\`{a} 
degli Studi and INFN, I-20133 Milano, Italy}, 
C.~Boiano\addressmark[INFNMI],
A.S.~Botvina\addressmark[GSI]\address[RUS]{Inst.~Nucl.~Res., 
Russian Accademy of Science, Ru-117312 Moscow, Russia},
A.~Boudard\address[FRA2]{DAPNIA/SPhN, CEA/Saclay, F-91191 Gif-sur-Yvette, France}, 
J.~Brzychczyk\addressmark[POL2], 
A.~Chbihi\address[GAN]{GANIL, CEA et IN2P3-CNRS, F-14076 Caen, France},
J.~Cibor\address[POL3]{H. Niewodnicza{\'n}ski Institute 
of Nuclear Physics, Pl-31342 Krak{\'o}w,Poland}, 
B.~Czech\addressmark[POL3],
M.~De~Napoli\addressmark[INFNCT], 
J.-E.~Ducret\addressmark[FRA2], 
H.~Emling\addressmark[GSI], 
J.~Frankland\addressmark[GAN], 
M.~Hellstr{\"o}m\addressmark[GSI], 
D.~Henzlova\addressmark[GSI], 
K.~Kezzar\addressmark[GSI],
G.~Imm{\'e}\addressmark[INFNCT], 
I.~Iori\addressmark[INFNMI], 
H.~Johansson\addressmark[GSI], 
A.~Lafriakh\addressmark[FRA1], 
A.~Le~F{\`e}vre\addressmark[GSI], 
E.~Le~Gentil\addressmark[FRA2], 
Y.~Leifels\addressmark[GSI], 
W.G.~Lynch\address[MSU]{Department of Physics and
Astronomy and NSCL, MSU, East Lansing, MI 48824, USA}, 
J.~L{\"u}hning\addressmark[GSI], 
J.~{{\L}}ukasik\addressmark[GSI]\addressmark[POL3], 
U.~Lynen\addressmark[GSI], 
Z.~Majka\addressmark[POL2], 
M.~Mocko\addressmark[MSU], 
W.F.J.~M{\"u}ller\addressmark[GSI], 
A.~Mykulyak\address[POL1]{A.~So{\l}tan Institute for Nuclear Studies,
Pl-00681 Warsaw, Poland}, 
H.~Orth\addressmark[GSI], 
A.N.~Otte\addressmark[GSI], 
R.~Palit\addressmark[GSI], 
A.~Pullia\addressmark[INFNMI], 
G.~Raciti\addressmark[INFNCT], 
E.~Rapisarda\addressmark[INFNCT], 
H.~Sann\addressmark[GSI]\thanks{deceased}, 
C.~Schwarz\addressmark[GSI], 
H.~Simon\addressmark[GSI],
A.~Sokolov\addressmark[GSI], 
K.~S{\"u}mmerer\addressmark[GSI], 
W.~Trautmann\addressmark[GSI],
M.B.~Tsang\addressmark[MSU],  
G.~Verde\addressmark[MSU], 
C.~Volant\addressmark[FRA2], 
M.~Wallace\addressmark[MSU], 
H.~Weick\addressmark[GSI],
J.~Wiechula\addressmark[GSI], 
A.~Wieloch\addressmark[POL2] and 
B.~Zwieglinski\addressmark[POL1] 
}
       
\begin{document}

\maketitle

\begin{abstract}
A systematic study of isospin effects in the breakup of projectile spectators
 at relativistic energies has been performed with the ALADiN spectrometer at 
the GSI laboratory (Darmstadt). Four different projectiles $^{197}$Au, 
$^{124}$La, $^{124}$Sn and $^{107}$Sn, all with an incident energy of 600~AMeV, have 
been used, thus allowing a study of various combinations of masses and $N/Z$ 
ratios in the entrance channel. \\
The measurement of the momentum vector and of the charge of all 
projectile fragments with $Z>1$ entering the acceptance of the ALADiN 
magnet has been performed with the high efficiency and resolution 
achieved with the TP-MUSIC IV detector. \\
The Rise and Fall behavior of the mean multiplicity of IMFs as a function 
of $Z_{{\rm bound}}$ and its dependence on the isotopic composition has been 
determined for the studied systems. Other observables investigated so 
far include 
mean $N/Z$ values of the emitted light fragments and neutron multiplicities.
Qualitative agreement has been obtained between the observed gross 
properties and the predictions of the Statistical Multifragmentation Model.
\end{abstract}

\section{Introduction}

The role of isospin in multifragmentation has been explored very little up
to now. This is in striking contrast to the importance of isospin, in
particular for any interpretation of multifragmentation as a manifestation
of the liquid-gas phase transition in nuclear matter. M{\"u}ller and Serot, 
in their seminal paper \cite{muell95}, have demonstrated that the two-fluid
nature of nuclear matter has very specific consequences for the phase 
behavior in the coexistence region. Different isotopic compositions are 
predicted for the coexisting liquid and gas phases, with the gas being more 
neutron rich than the liquid in asymmetric matter.
This difference stems from the decrease in the symmetry energy in nuclear 
matter as the density is decreased. The expected magnitude of this density 
dependence, however, is model dependent and very poorly constrained by 
existing data \cite{bomb91}.\\
Furthermore, the calculations are restricted to infinite matter
with neither Coulomb forces nor fragment formation included. 
In addition, the assumed isotopic composition 
is typically varied within a range of proton fractions whose limits are 
not easily accessible 
in experiments with heavy nuclei. Theoretical studies for finite 
systems also indicate that the sequential decay of excited reaction 
products has a tendency to modify some of the expected effects \cite{lari99}.
Some observables are, however, predicted to be robust with respect 
to sequential decay. If chemical equilibrium can be assumed, isotopic ratios
can be constructed whose exponential dependence on the 
ratio $\mu /T$ of the chemical potential $\mu$ 
and the temperature $T$ will amplify differences resulting from the 
variation of $\mu$ with the isotopic composition. 
Measured isotopic yield ratios were found to
vary strongly with the $N/Z$ ratio of the emitting source \cite{wada87},
in agreement with this expectation \cite{barz88,hahn88}.
Recently, significant differences between 
neutron-rich and neutron-poor systems have been observed~\cite{xu00}. 
In particular, the strong enhancement of the neutron content in the 'gas' 
with increasing $N/Z$ of the system has been interpreted as being 
in qualitative agreement with the 
predictions of \cite{muell95}. The role of the excitation energy
has also been emphasized \cite{mila00} and evidence has been found for 
the increased production of neutron-rich isotopes with excitation energy
that is predicted by the Statistical Multifragmentation 
Model (SMM,~ref.~\cite{bond95}).

\begin{figure}[ttb]
\centering
\begin{minipage}[c]{.49\textwidth}
   \centerline{\includegraphics[height=7.5cm]{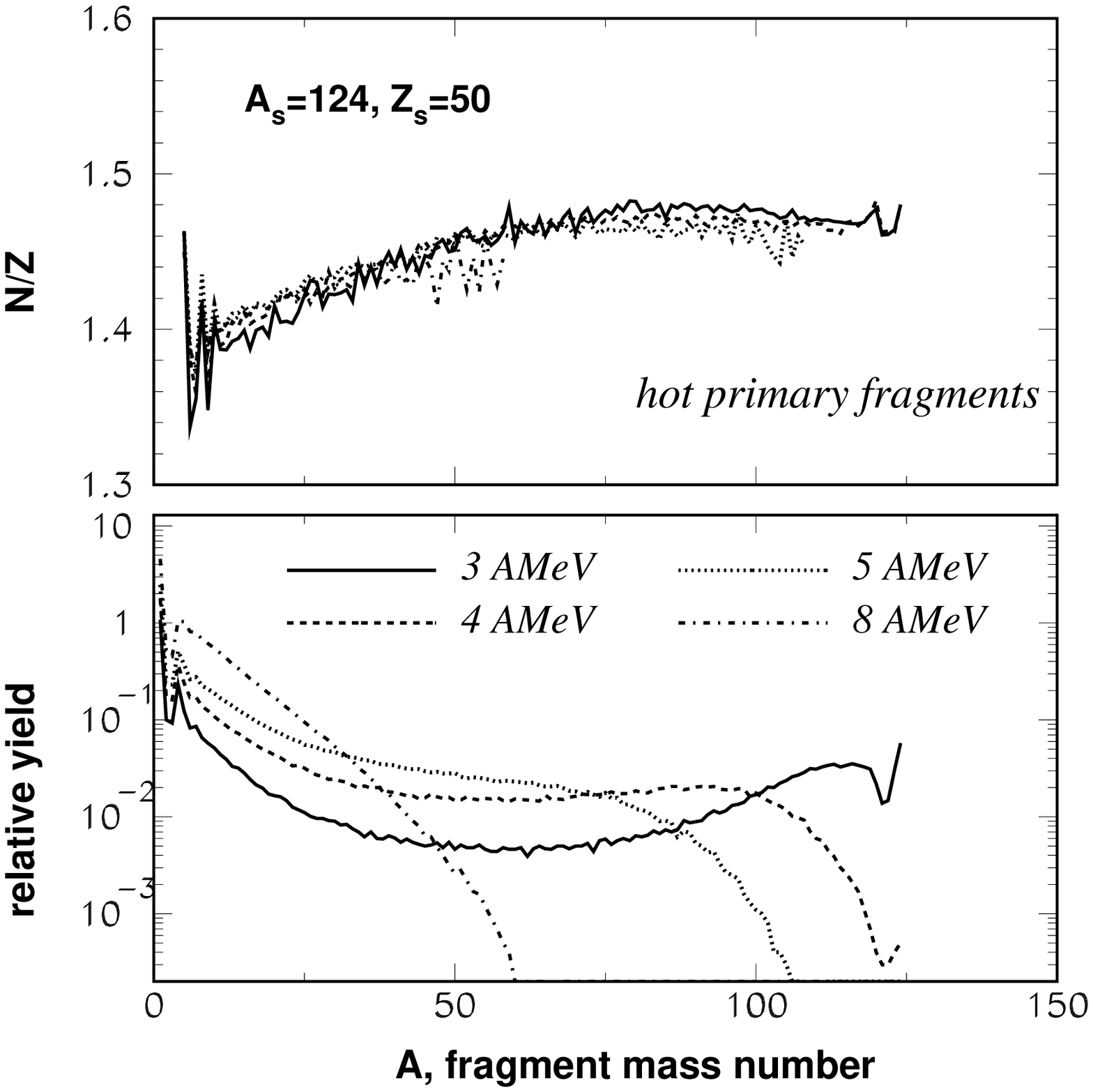}}
\end{minipage}
\begin{minipage}[c]{.49\textwidth}
   \centerline{\includegraphics[height=7.5cm]{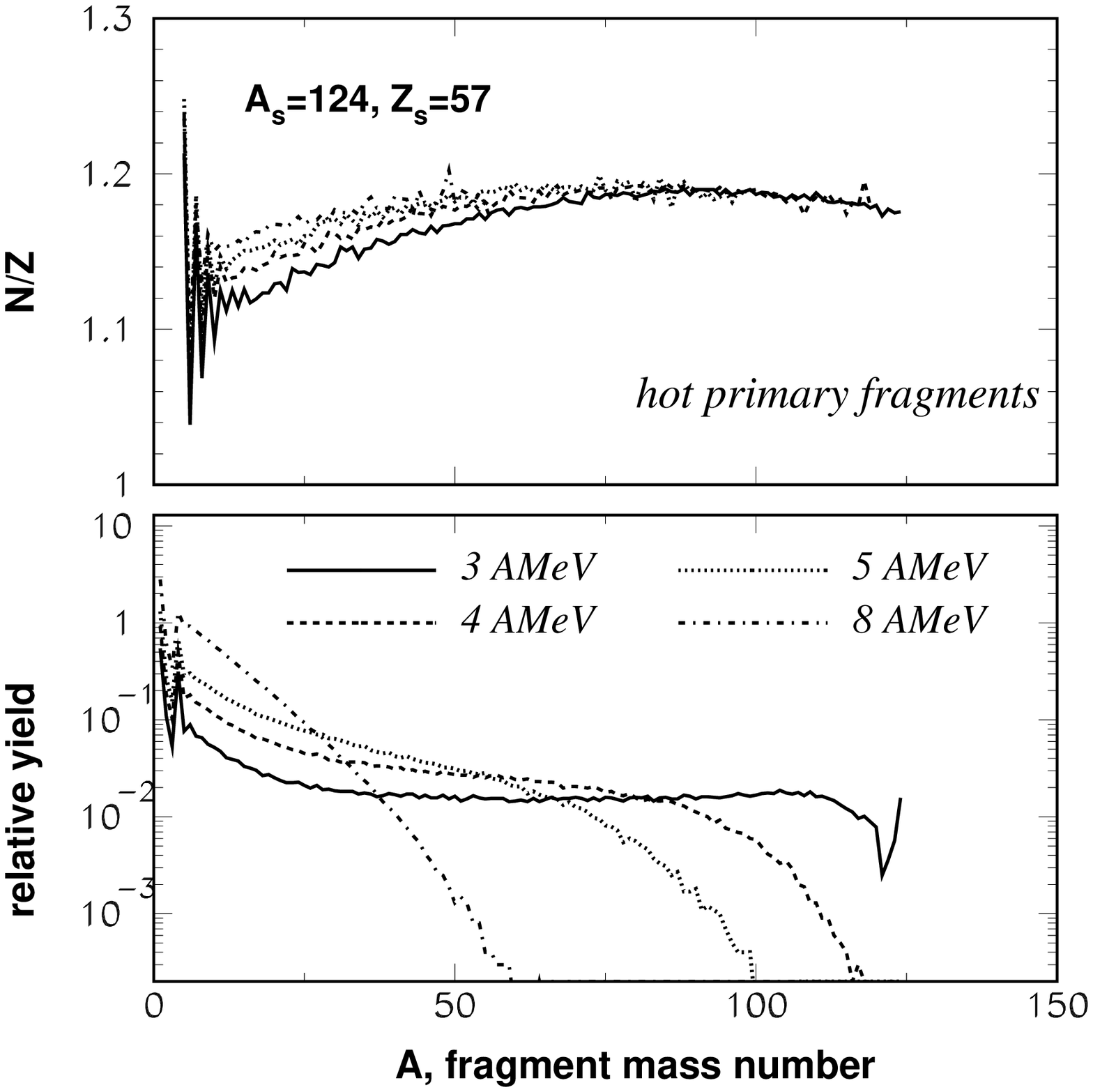}}
\end{minipage}
   \caption{\it\small 
Neutron-to-proton ratios $N/Z$ (top panels)
and mass-yield curves (bottom panels) of the primary hot 
fragments produced at the breakup of two systems with fixed mass
number $A$~=~124 according to SMM calculations. The lines correspond to
four excitation energies between 3 and 8 MeV per nucleon as indicated.
Note the difference of the ordinate scales in the top panels
(from Ref. \protect\cite{botv00}).
}
\label{smm}
\end{figure}

The SMM predictions for the fragmentation of two
$A$ = 124 systems, one neutron-rich ($N/Z$ = 1.48) and one neutron-poor 
($N/Z$ = 1.18), are given in Fig.~\ref{smm}.
The isotopic compositions of the hot fragments produced 
at breakup are globally very different for the two systems, close
to the $N/Z$ ratios of the primary projectiles which they approach
with increasing mass number (top panels). 
The overall mass dependence is rather weak, with a dependence on 
excitation energy predicted to be different for the neutron-rich and 
neutron-poor cases. 
The calculated mass yield curves are fairly similar for these systems 
except at low excitation energy (bottom panels of Fig. \ref{smm}). 
Exploring these dependences
experimentally will be important for our understanding of the role of phase 
space in the multifragment decays.\\ 
\begin{figure}[htb]
   \centerline{\includegraphics[height=8cm,angle=-90]{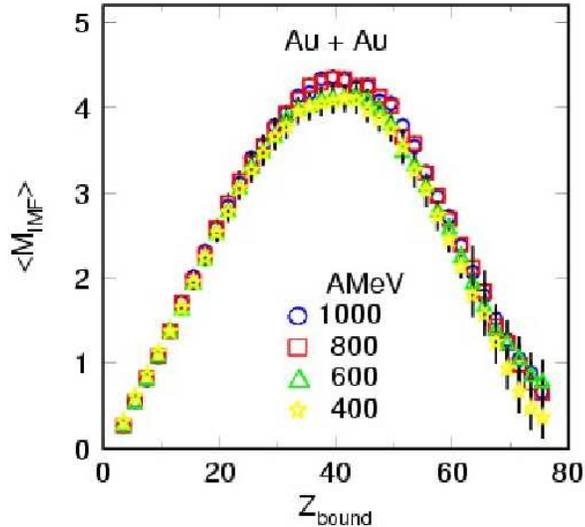}}
        \caption[]{\it\small
Rise and Fall of multifragment emission \cite{schuett96}: the mean
multiplicity of intermediate mass fragment is plotted as a function of 
$Z_{{\rm bound}}$ for different energies.
}
\label{riseandfall}
\end{figure}
In a series of experiments \cite{schuett96,poch95}, multifragment decay 
of projectile spectators has been studied with the ALADiN forward-spectrometer
 at the SIS accelerator (GSI). The primary projectile spectator emerging 
from the abrasion phase is a well defined source which can be reconstructed 
on an event-by-event basis from the multiplicity and energy of the 
fragments. The isotropy of the fragment emission in the decaying spectator 
rest-frame suggests an emission from a thermodynamical equilibrated source. 
In these collisions, energy depositions are reached, which cover the range 
from particle evaporation to multifragment emission and further to the 
total disassembly of the system, the so-called {\em Rise and Fall of 
multifragment emission} \cite{schuett96} (Fig.~\ref{riseandfall}). 
The most prominent feature of the multi-fragment decay is the universality 
of the fragment multiplicites and the fragment charge correlations. 
These observables are invariant with respect to the entrance channel, i.e. 
independent of the beam energy and the target or projectile masses, 
if plotted as a function of $Z_{{\rm bound}}$. This latter quantity, being the sum 
of the atomic numbers $Z_{\rm i}$ of all projectile fragments with $Z_{\rm i} > 1$,
represents a measure of the size of the spectator and, in a geometrical 
picture, is monotonously correlated with the impact parameter.
The loss of memory of the entrance channel is an indication that 
equilibrium is attained prior to the fragmentation stage of the reaction.
It will be interesting to investigate whether the observed universality of 
spectator decays includes the invariance with isospin.\\ 
\begin{figure}[htb]
   \centerline{\includegraphics[height=10cm,angle=-90]{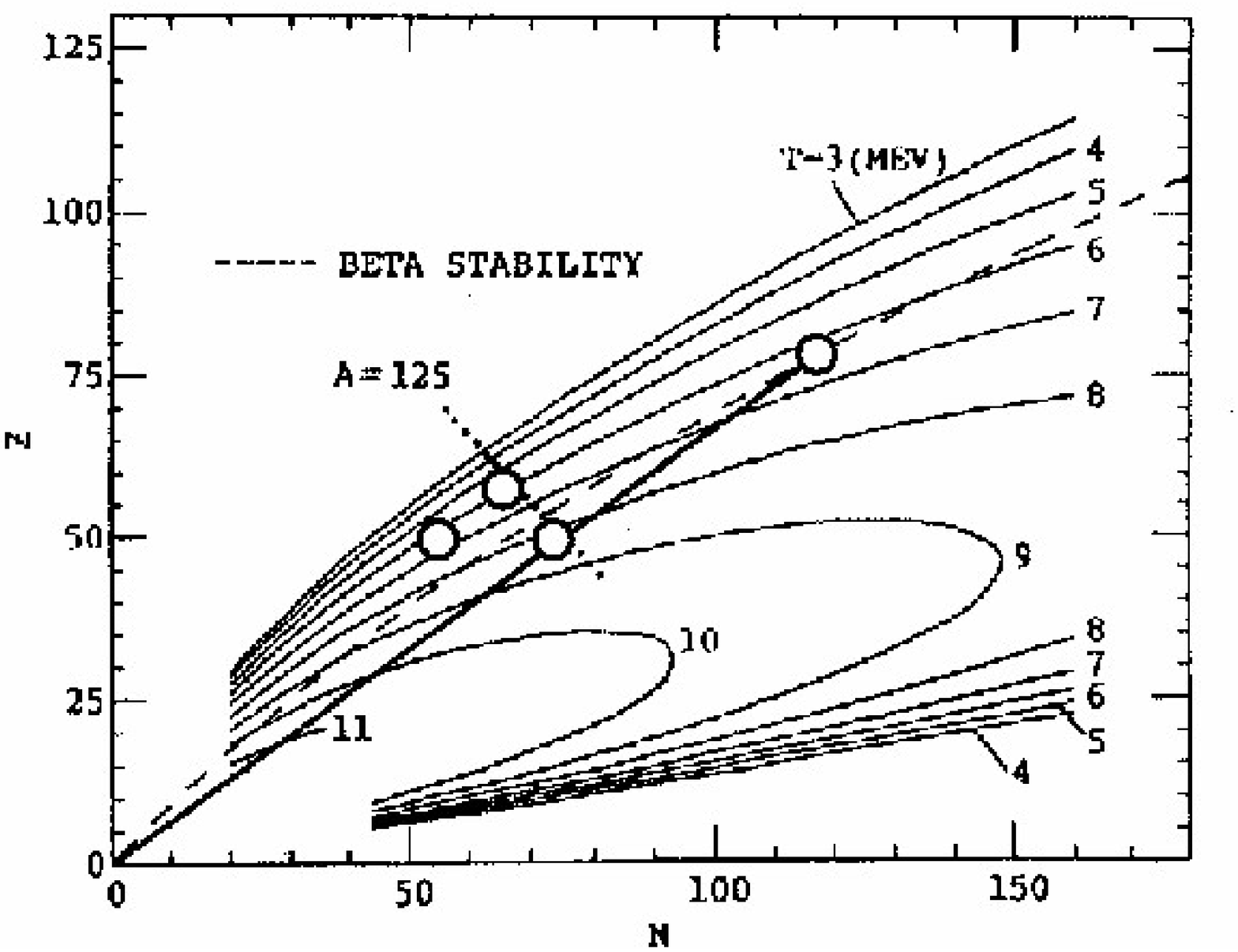}}
        \caption[]{\it\small
Location of the four studied projectiles in the plane of atomic number $Z$ 
versus neutron number $N$. The contour lines represent the limiting 
temperatures according to Ref. \protect\cite{besp89}, the dashed line 
gives the valley of stability, and the full line corresponds to the 
$N/Z$ = 1.49 of $^{197}$Au.
}
\label{limit}
\end{figure}
Besides the isospin, also the mass of the system may play an important role 
for fragmentation observables of the reaction. It has been suggested that 
the dependence of the breakup temperature on the excitation energy 
(caloric curve) is governed by the {\em limiting temperature} 
\cite{nato95,nato01}. 
This quantity 
represents the maximum temperature at which nuclei are found
to exist as self-bound objects in Hartree-Fock calculations \cite{besp89}.
In lighter systems the limiting temperature is higher, mainly so because 
the Coulomb energy is reduced (Fig.~\ref{limit}). 
On the other hand,
SMM calculations predict nearly mass-invariant temperatures for 
the coexistence region \cite{bond95}. 
The comparison of two systems with different mass should, therefore,
permit distinguishing whether the breakup temperature is
determined by the binding properties of 
the excited hot nuclear system or by the phase space accessible to it by 
fragmentation. The same test can be made by varying the isospin.

\section{The Experimental Setup}

The most recent ALADiN experiment has been devoted to investigating 
isotopic effects in the decay of projectile spectators at relativistic 
energies. 
In order to extend the range of isotopic compositions of the excited 
spectator systems, secondary beams have also been used.
This and the clean separation of the spectator sources in rapidity
make this type of reaction unique for studying the isospin dependence of
nuclear multifragmentation.\\
\begin{figure}[htb]
 
\centerline{\includegraphics[angle=-90,scale=1.3,
            bbllx=80,bblly=10,bburx=220,bbury=450]{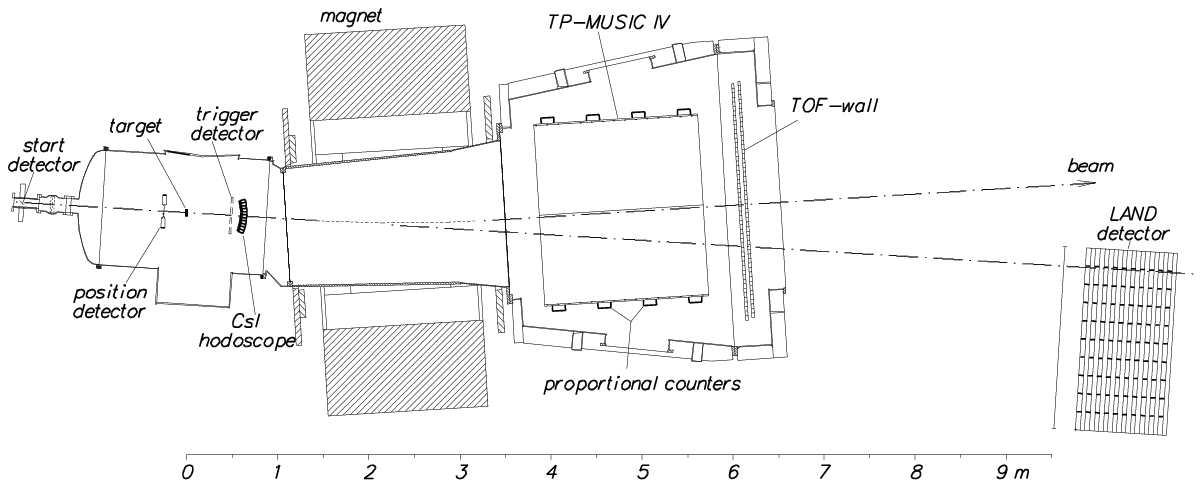}}
        \caption[]{\it\small
Cross sectional view of the ALADIN setup.
}
\label{setup}
\end{figure}
Four different projectiles, all with an incident energy of 600~AMeV, 
have been investigated allowing a study of various combinations of 
masses and $N/Z$ ratios in the entrance channel: 
$^{124}$Sn, $^{197}$Au, $^{124}$La and $^{107}$Sn. The two 
latter beams have been delivered by the FRagment Separator (FRS) of the
GSI as products of the fragmentation of a primary $^{142}$Nd beam at 1.1~AGeV 
on a $^9$Be production target. 
The necessity of low beam intensities for the best operational 
condition of the ALADiN setup ($\approx$ 2000 particles/sec), and the possibility 
of using a thick target
in order to achieve high interaction rates are indeed conditions
compatible with radioactive-ion-beam experiments.
Moreover, the inverse kinematics offers the possibility of a
threshold-free detection of all heavy fragments and residues and thus
gives a unique access to the breakup dynamics.\\
A cross sectional view of the used setup is shown in Fig.~\ref{setup}.
The beam enters from the left and passes thin time- and position-detectors 
before reaching the  $^{\rm nat}$Sn target with $500$-mg/cm$^2$ areal density. 
The isotopic composition of the secondary beams 
was determined and monitored from the magnetic rigidity measured at the FRS, 
from a velocity measurement along the $80$-m flight path between the FRS 
and the ALADiN setup, and from the charge measurement with the TP-MUSIC IV detector.\\
Projectile fragments entering into the acceptance of the magnet are
tracked and identified in the TP-MUSIC IV detector and in the
time-of-flight (TOF) wall. 
Neutrons emitted in directions close to $\theta_{{\rm lab}} = 0^{\circ}$,
are detected with the Large-Area Neutron Detector (LAND).
The dash-dotted lines represent the beam directions before and after the 
deflection by 7$^{\circ}$ in the field of the ALADiN magnet. \\
The measurement of the charge and the momentum vector of 
all projectile fragments with $Z\geq2$ 
has been performed with high efficiency and 
high resolution with the TP-MUSIC IV detector. 
In order to cover the wide dynamic range necessary to measure nuclei from
He up to Au with the best possible resolution, two different kinds of 
detectors are connected to the field cage of the TP-MUSIC 
detector on either side~\cite{bauer}: 
the ionization charge collected at 24 anodes provides optimum Z resolution 
for heavy fragments ($Z>8$), whereas the 3D tracking information of all 
particles 
and the charges of lighter fragments are obtained from 4 position-sensitive 
proportional counters. The position of the ionizing particles 
in the non-bending plane is determined from the position 
along the proportional counters, whereas the position in the bending plane
is determined by measuring the total drift time of the electrons to the 
detectors.
\begin{figure}[htb]
   \centerline{\includegraphics[width=10cm,height=10cm]{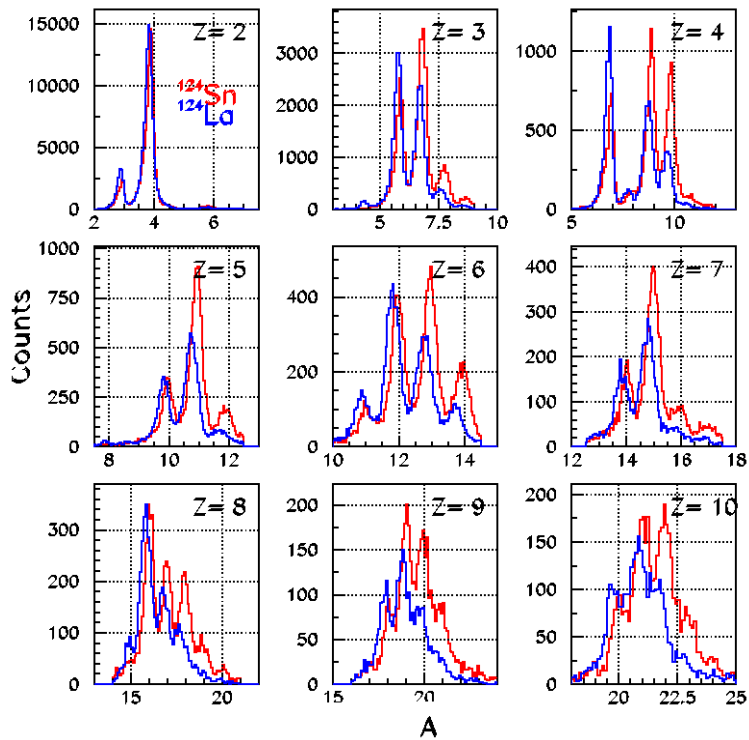}}

        \caption[]{(Color online) \it\small Mass spectra for light fragments with $Z \leq 10$ from 
the fragmentation of $^{124}$La (blue) and $^{124}$Sn (red).}
\label{resolution}
\end{figure}
In order to improve the performance of the TP-MUSIC detector, an upgrade has
been undertaken~\cite{jahr2002} which involved the construction of a new 
set of proportional counters and a redesign of all components of the 
electronic chain compared with the previous version of the 
detector~\cite{bauer}.
The proportional counters use a combination of charge-division and pad-readout 
techniques to reconstruct the position in the non-bending plane of the tracks of nuclei~\cite{jahr2002}. 
The pads of each section are connected modulo five: the resulting position ambiguity can indeed be resolved by using 
the less precise position information obtained from the anode wires 
with the charge-division method.\\
To extract the signals from each sector of the proportional counters, seven 
charge-sensitive preamplifiers are used.
The signals of the preamplifiers, after removal of the high-frequency 
component through an anti-aliasing filter, are digitized by 14-bit Flash ADC's
without prior shaping. The typical noise is of 1 Least Significant Bit 
(standard deviation) over a dynamic range of $1:10^4$.
The output, generated at a rate of up to 40 MHz, is stored and processed 
by a system containing FPGA and DSP chips.\\
Using the reconstructed values for the rigidity and path length, 
the charge of the particle measured by the TP-MUSIC detector, and 
the time of flight given by the TOF-wall,
the velocity and the momentum vector can be calculated for each detected 
charged particle.
The knowledge of velocity and momentum allows then the calculation of the 
particle's mass.
Mass spectra for the $^{124}$Sn and $^{124}$La fragmentation are shown 
in Fig.~\ref{resolution}. 
Single mass resolution for charges up to 12 is obtained, corresponding to
a mass resolution $\Delta A/A$ of approximately 4.0\% (FWHM) for light 
fragments. \\
Already by inspection of the mass spectra, 
first order isotopic effects can be observed: in the case 
of the neutron-rich system higher yields 
are obtained for the neutron-rich isotopes.

\section{Gross Properties of the multifragment decay}

\begin{figure}[ttb]
\centering
\begin{minipage}[c]{.49\textwidth}
   \centerline{\includegraphics[height=8.5cm]{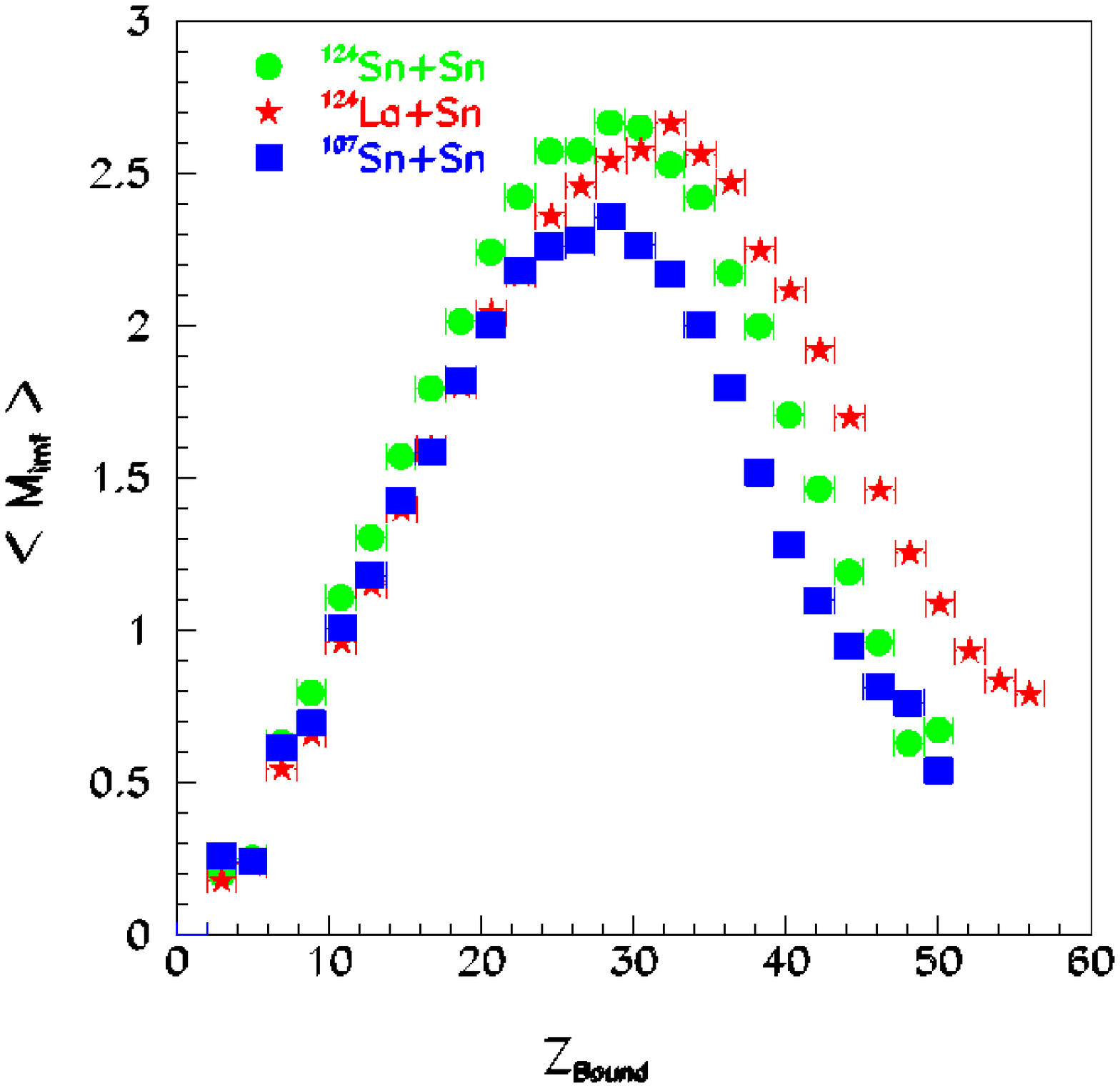}}
\end{minipage}
\begin{minipage}[c]{.49\textwidth}
   \centerline{\includegraphics[height=7cm]{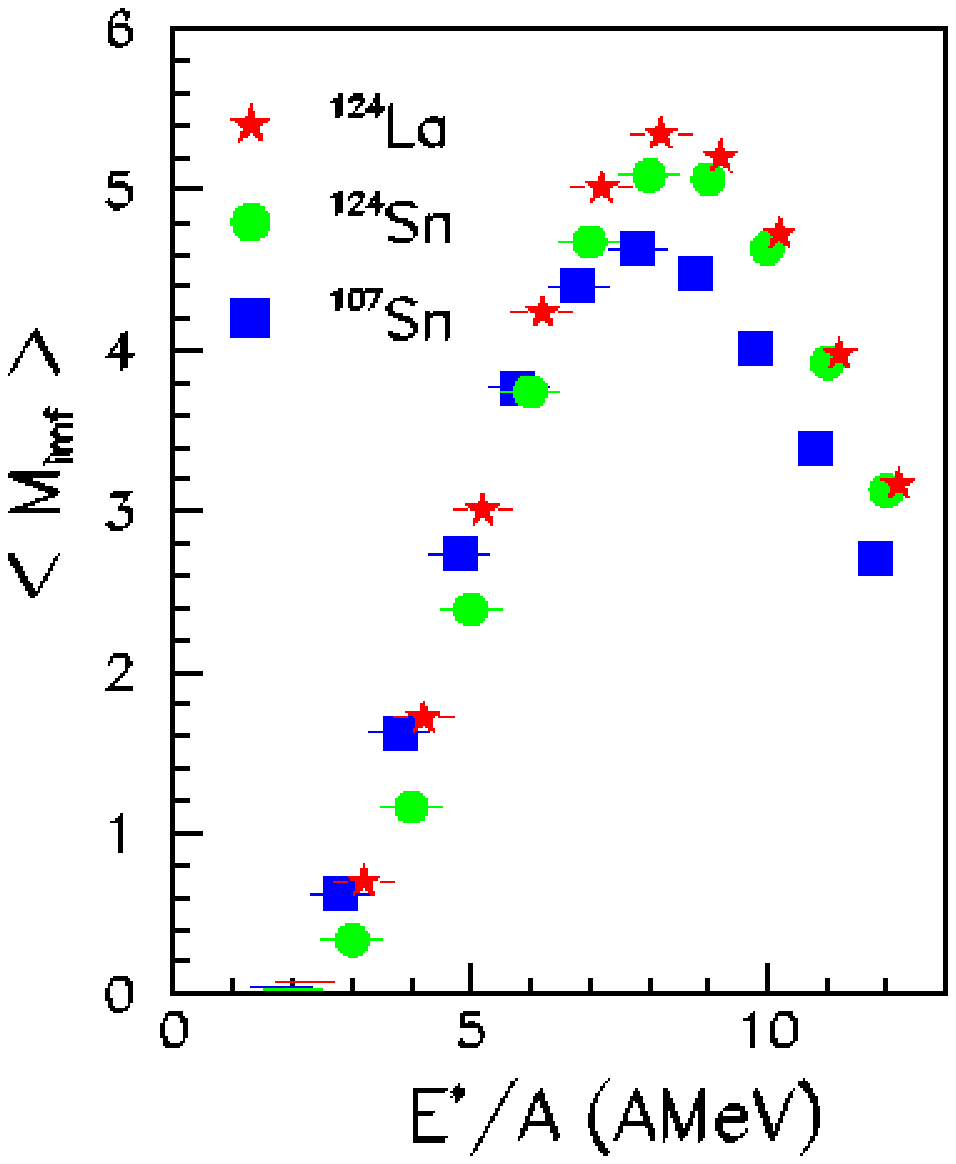}}
\end{minipage}
   \caption{\it\small Left Panel: Experimental Rise and Fall of multifragmentation correlating the 
   mean multiplicity of intermediate-mass fragments and $Z_{{\rm bound}}$. 
   Right Panel: Prediction of the Statistical Multifragmentation Model for 
   the mean IMF multiplicity as a function of the excitation energy for excited 
   $^{124}$La, $^{124}$Sn and $^{107}$Sn nuclei.
}
\label{grossprop}
\end{figure}

\begin{figure}[ttb]
\centering
\begin{minipage}[c]{.49\textwidth}
   \centerline{\includegraphics[height=7.6cm]{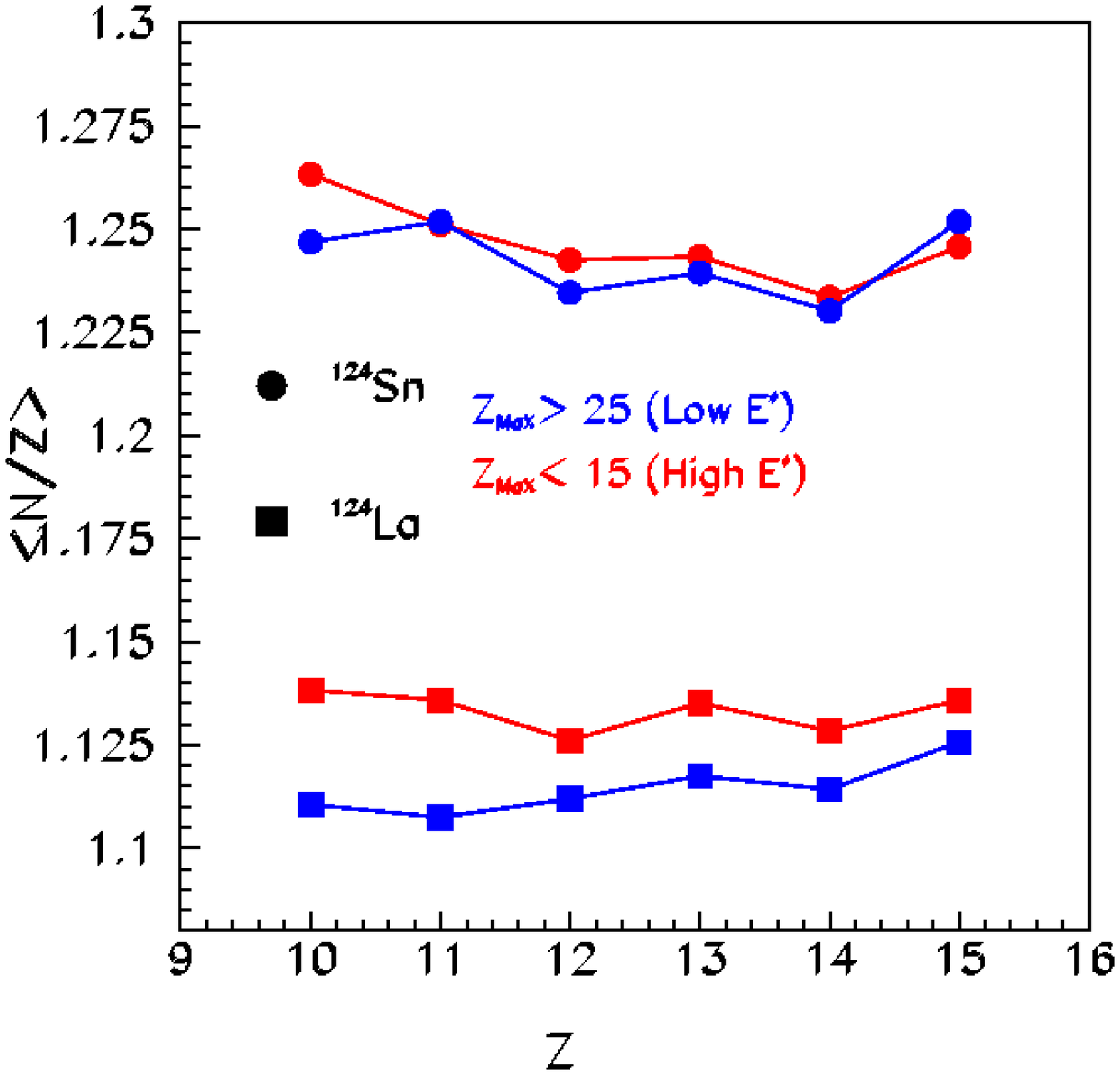}}
\end{minipage}
\begin{minipage}[c]{.49\textwidth}
   \centerline{\includegraphics[height=7.6cm]{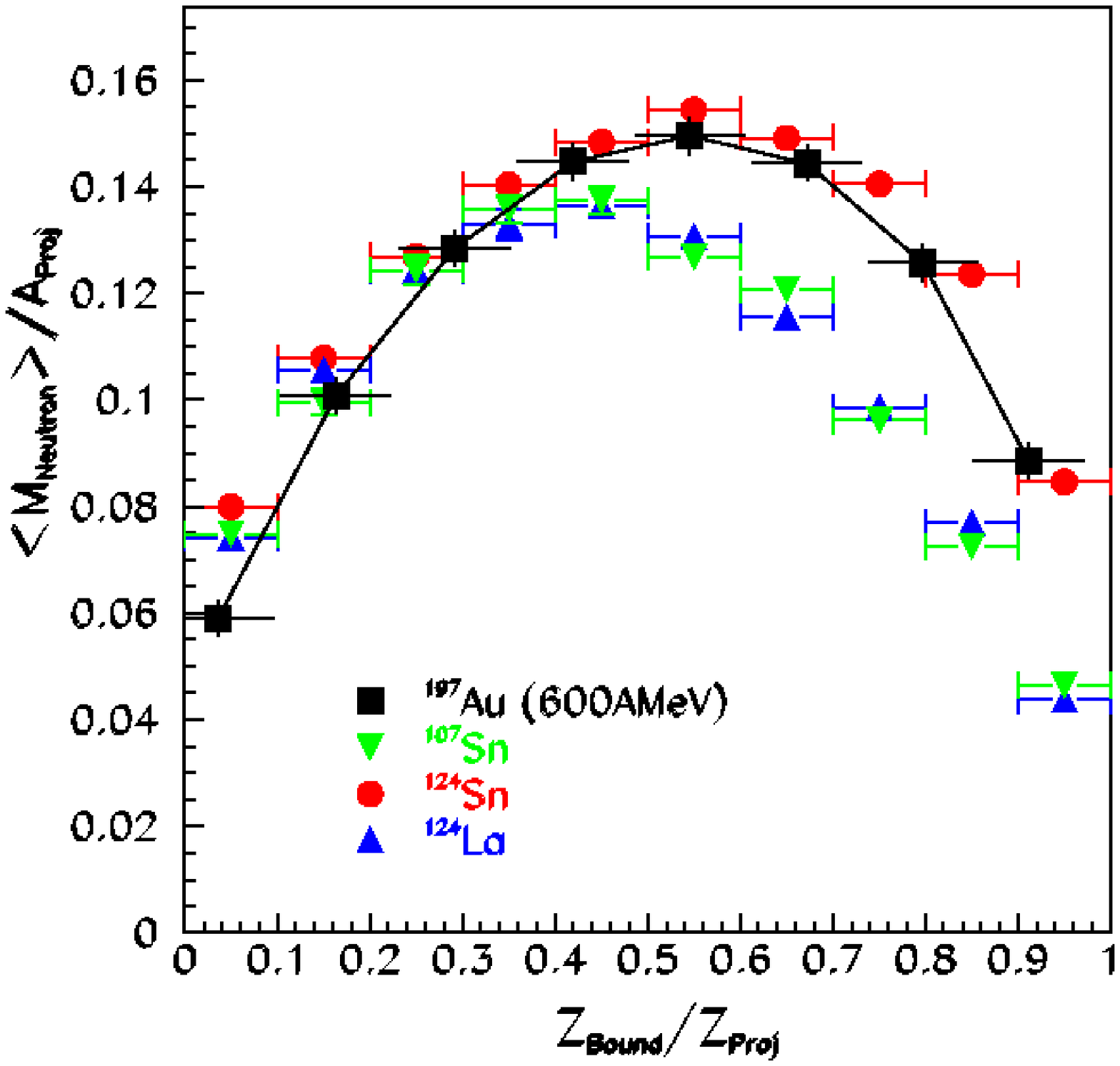}}
\end{minipage}
   \caption{\it\small Left Panel: Mean $N/Z$ distribution for medium-size 
fragments for two cuts in the variable $Z_{{\rm max}}$ (corresponding to two 
different excitation energy intervals). Right Panel: Preliminary mean neutron 
multiplicity, as derived from the raw hit multiplicity in Land, 
corrected for solid angles and scaled for projectile mass as a function of 
$Z_{{\rm bound}}$ for the three analyzed systems compared with the data for 
$^{197}$Au projectile fragmentation~\cite{zude}.
}
\label{nzneutrons}
\end{figure}

In order to investigate to which extent the isotopic composition of the 
excited spectator affects the gross properties of the multifragmentation 
pattern, charge partitions and multiplicity distributions have been 
analyzed, as well as the mean $N/Z$ of medium size fragments, and the results
have been compared with the SMM prediction.\\
In Fig.~\ref{grossprop}, the obtained correlation between the mean 
multiplicity of intermediate-mass fragments, $<M_{IMF}>$, and the variable 
$Z_{{\rm bound}}$ for the $^{107}$Sn, 
$^{124}$Sn and $^{124}$La systems is shown (left panel).
The global universality of the {\rm Rise and Fall} behavior 
is preserved, but already some distinct differences can be observed. 
At small excitation 
energies (large $Z_{{\rm bound}}$ values) the curves end, as expected, 
approximately at the 
charge of the original projectiles. However, the slope of the curve is
steeper in the case of the $^{124}Sn$.
This effect can be understood by considering that in the
case of the neutron-rich system, heavy residues with low excitation energy 
will predominantly emit neutrons, a channel that is suppressed 
in the case of the two neutron-poor nuclei. In these latter cases, peripheral collisions are more spread out towards
smaller values of $Z_{{\rm bound}}$, thus leading to a slower rise of $<M_{IMF}>$. 
This effect, as well as the corresponding difference in the $Z_{\rm{bound}}$ 
distribution, is in good agreement with the SMM predictions.\\
Going towards more central collisions, we observe a lower maximum in $<M_{IMF}>$ 
for the lighter Sn isotope, whereas the two A=124 systems exhibit the same
value for the mean multiplicities. 
This mass effect, reproduced by SMM calculations 
(Fig.~\ref{grossprop} right panel) is rather intriguing since 
the definition of IMF ($3\leq Z \leq 20$) is based on the charge. 
Therefore, the partition space should be primarily a function of the charge 
of the source.
In the rise, up to 6 or 7 MeV per nucleon, the number of fragments is just 
given by the excitation energy. The $^{107}$Sn bends over more quickly towards 
vaporization: less neutrons are available and we may have more alpha-type 
fragments (like $^8$Be, $^{12}$C) that decay easily into alpha particles. 
The absolute multiplicities are overpredicted because the calculations are 
performed for sources of fixed mass and not for the ensemble of spectator
systems produced in the collision.
Very specific isotopic effects, even though small, can also be 
extracted from
the analysis of the mean $N/Z$s for medium-size fragments. 
In this case, in order to generate two data samples selected by 
excitation energy, two cuts on the 
maximum charge detected in each event ($Z_{{\rm max}}$) have been used: 
the data have been correspondingly sorted into two different bins of the
 variable $Z_{{\rm max}}$, corresponding to high and low excitation energies. 
The mean $N/Z$s for medium-size fragments obtained are shown in 
Fig.~\ref{nzneutrons}. For the $^{124}$Sn, the values are always bigger 
than the ones 
obtained in the case of the $^{124}$La: this observation reflects simply 
the difference in the $N/Z$ of the original projectiles. 
In the case of the $^{124}$La, on the other hand, 
a difference in the mean $N/Z$s 
corresponding to the two cuts is observed, whereas the $^{124}$Sn shows almost 
no sensitivity. 
Also this characteristics is predicted by the SMM calculations as evident 
from Fig.~\ref{smm}.
This latter case, however, refers to hot primary fragments whereas the 
extracted $<N/Z>$ correspond to the  
detected final fragments: this explains why the absolute 
scale in the data is much lower than the predicted one.  It is, however, 
interesting to see that this feature seems to survive the sequential decays.\\
In the statistical scenario, this difference arises from the dependence of the
number of neutrons which will have to be carried by the light fragments on 
the yields of heavy fragments and their capability of carrying neutrons 
(depends on mass and $N/Z$).
This effect supports the idea of overall equilibrium of the system at breakup.

\section{Neutron Multiplicities}

As already mentioned, the used experimental setup allowed also 
neutron detection.
In Fig.~\ref{nzneutrons} the mean neutron multiplicity, corrected 
for solid angle and scaled for projectile mass, has been correlated with the 
variable $Z_{{\rm bound}}$.
A very clear isotopic effect (again first-order effect) is visible: more 
neutrons are produced in the case of the neutron-rich system. In particular 
a good agreement has been found by comparing the scaled neutron multiplicity 
measured 
for the $^{124}$Sn system with the one obtained in the case of the heavier 
$^{197}$Au projectile~\cite{zude}, which has the same $N/Z$.
This difference decreases with excitation energy and the curves 
nearly merge.
Neutrons will be important for establishing the mass and energy balance, 
in particular for calorimetry. In this respect, it is crucial to identify 
the spectator neutrons and to distinguish them from the fireball ones.\\
Moreover in the grand-canonical approximation it can be demonstrated that,
from the ratio of the neutron yields the symmetry term of the 
nuclear equation of state can be determined once the temperature and 
the isotopic composition of the systems are known~\cite{traut}.
In this respect, neutron analysis could allow to investigate the
symmetry-term dependence on the excitation energy of the system, in a 
similar way as with the isoscaling analysis~\cite{lef2004}.

\section{Conclusions}
First preliminary results from the most recent ALADiN experiment devoted to 
the investigation of mass and isospin effects in multifragmentation have been
reported. Global quantities as, e.g., the mean multiplicity of 
intermediate-mass fragments exhibit small but significant variations 
with the isotopic composition of the fragmenting projectile. \\
Much larger effects are observed 
for the $N/Z$ ratios of the emitted fragments. Their characteristic 
dependence on excitation energy promises to allow a meaningful test of the 
attainment of statistical equilibrium.\\
The measured neutron yields have a maximum at intermediate values of 
${Z_{\rm bound}}$, practically coinciding with the maximum fragment multiplicity. 
Their isotopic dependence is large for large ${Z_{\rm bound}}$ but gradually 
disappears with decreasing $Z_{{\rm bound}}$, i.e. for the highest excitation 
energies.\\
{\it The authors would like to thank the staff of the GSI for 
providing heavy ion beams of the highest quality and for technical support.
C.Sf. acknowledges the receipt of an Alexander-von-Humboldt fellowship.
This work was supported by the European Community under
contract No. HPRI-CT-1999-00001 and by the Polish Scientific Research 
Committee under contract No. 2P03B11023.}

\end{document}